\documentclass[useAMS,usenatbib]{mn2e}

\usepackage {graphicx}
\usepackage {times}

\title[A Catalog of Local E+A(post-starburst) Galaxies selected from the Sloan Digital Sky Survey Data Release 5]{A Catalog of Local E+A(post-starburst) Galaxies selected from the Sloan Digital Sky Survey Data Release 5}
\author[T. Goto]{Tomotsugu Goto
 \thanks{E-mail:tomo@ir.isas.jaxa.jp} 
\\
 Institute of Space and Astronautical Science,  Japan Aerospace Exploration Agency,
 3-1-1 Yoshinodai, Sagamihara, Kanagawa 229-8510, Japan}
\begin{document}

%\date{Accepted 2006 December 15. Received 1988 December 14; in original form 2006 March 17}
\date{\today; in original form 2007 February 1}

\pagerange{\pageref{firstpage}--\pageref{lastpage}} \pubyear{2006}

\maketitle

\label{firstpage}

\begin{abstract}
 E+A galaxies have been interpreted as post-starburst galaxies based on
 the presence of strong Balmer absorption lines combined with the
 absence of major emission lines ([OII] nor H$\alpha$). As a population of galaxies in the midst of the transformation, E+A galaxies has been subject to an intense research activity. It has been, however, difficult to investigate E+A galaxies statistically since E+A galaxies are an extremely rare population of galaxies ($<$1\% of all galaxies in the local Universe).

 Here, we present a large catalog of 564 E+A (post-starburst) galaxies carefully
 selected from half million spectra of the Sloan Digital Sky Survey Data
 Release 5. We define E+A galaxies as those with H$\delta$ equivalent
 width $>$5\AA\  and no detectable emission in [OII] $and$ H$\alpha$.
 The catalog contains 564 E+A galaxies, and thus, is one of the largest of the
 kind to date. In addition, we have included the H$\alpha$ line in the  selection to remove dusty star-forming galaxies, which could have contaminated previous [OII]-based samples of E+A galaxies up to 52\%. 
  Thus, the catalog is one of the most homogeneous, let alone its size.
 The catalog presented here can be used for follow-up observations and statistical analyses of this rare population of galaxies.

\end{abstract}


\section{Introduction}

%\subsection{post-starburst  interpretation of E+A galaxies}
%Ｅ＋Ａ銀河はポストスターバーストである。 と解釈されて注目を浴びてきた。
\citet{1983ApJ...270....7D,1992ApJS...78....1D} found galaxies with mysterious spectra while investigating high redshift cluster galaxies.
 The galaxies had strong Balmer absorption lines with no
 emission in [OII]. These galaxies are called  ``E+A''
 galaxies since their spectra looked like a superposition of that of
 elliptical galaxies (Mg$_{5175}$, Fe$_{5270}$ and Ca$_{3934,3468}$
 absorption lines) and that of A-type stars (strong Balmer absorption lines)
\footnote{Since some of E+A galaxies are
 found to have disk-like morphology \citep{1994ApJ...430..121C,1994ApJ...430..107D,1997AJ....113..492C,1999ApJS..122...51D},
 these galaxies are sometimes called 
 ``K+A'' galaxies.  However, \citet{2003PhDT.........2G} found that their E+A sample with higher completeness has early-type morphology. Following this discovery, we call
 them as ``E+A'' throughout this work.}.
  The existence of strong Balmer absorption lines shows 
 that these galaxies have experienced starburst recently \citep[within a gigayear;][]{2004A&A...427..125G}.  However, these galaxies do not show any sign of on-going star
 formation as non-detection in the [OII] emission line
 indicates.  
   Therefore, E+A galaxies have been interpreted as a post-starburst galaxy,
 that is,  a galaxy which truncated starburst suddenly \citep{1983ApJ...270....7D,1992ApJS...78....1D,1987MNRAS.229..423C,1988MNRAS.230..249M,1990ApJ...350..585N,1991ApJ...381...33F,1996ApJ...471..694A,1999ApJS..122...51D,1999ApJ...518..576P,2004ApJ...617..867D,2003PASJ...55..771G,2004A&A...427..125G,2005MNRAS.357..937G,goto2006}.  
 However, the reason why they started starburst, and why they abruptly stopped starburst remain one of the mysteries in galaxy evolution.

 One of the major difficulties in investigating E+A galaxies has been their rarity: less than 1\% galaxies are in the E+A phase in the present Universe \citep{2003PASJ...55..771G}. 
 In this work, we solve this problem by creating a large catalog of 564 E+A galaxies carefully selected from the fifth public data release of the Sloan Digital Sky Survey \citep[SDSS;][]{2006ApJS..162...38A}.
  In addition to its size, due to the availability of the information on the H$\alpha$ line, this  catalog is one of the most homogeneous of the kind, nicely removing the contanimation from the H$\alpha$ emitting galaxies. 
 The catalog is publicly released\footnote{ http://www.ir.isas.jaxa.jp/$^{\sim}$tomo/ea5} so that world-wide researchers can utilize for statistical analyses and follow-up observations in various wavelength.
   Unless otherwise stated, we adopt the best-fit WMAP cosmology: $(h,\Omega_m,\Omega_L) = (0.71,0.27,0.73)$ \citep{2003ApJS..148....1B}.

%-Blake et al.
%-Tran et al.2003,2004
%-Liu et al.
%- redshift histogram. recommend z>0.05 cut. aperture bias plot.
%- Hdelta histogram. 
%- fraction as a function of z
%- aperture bias
%- pretty pictures
%- hd vs ha

\section{Catalog}\label{catalog}

\begin{figure}
\begin{center}
\includegraphics[scale=0.5]{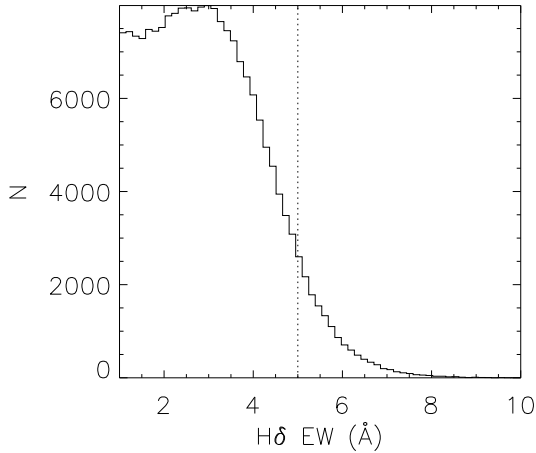}
\end{center}
\caption{
A histogram of the H$\delta$ equivalent width (\AA) of all galaxies. We select those
 with H$\delta$ EW $>5$\AA\  as E+A galaxies. 
}\label{fig:hd_hist}
\end{figure}

\subsection{Selection}

 Our new catalog of 564 E+A galaxies is based on the fifth public data release of the SDSS \citep{2006ApJS..162...38A}. This is the final data release of the SDSS I , and containts spectra of 675,000 galaxies. The selection algorithm is similar to that in \citet{2005MNRAS.357..937G}, which we summarize below \citep[also used in][]{2004A&A...427..125G,2005MNRAS.359.1557Y,2006AJ....131.2050Y,2006ApJ...642..152Y}.

  We only use those objects classified as galaxies ({\tt type=3}, i.e.,
 extended objects) with spectroscopically classified not to be a star
 ({\ttfamily SpecClass} is not a star) and the spectroscopic signal-to-noise $>$10 per pixel (in the continuum of the $r$-band wavelength range). These criteria nicely remove contamination from nearby stars and star-forming regions. 
 For these galaxies, we have measured H$\delta$,
  [OII] and H$\alpha$ equivalent widths (EWs) and obtained their errors
  using the flux summing  method described in \citet{2003PASJ...55..771G}. 
  This flux summing technique is advantageous over the Guassian fitting method in measuring reliable line strengths for noisy spectra. % or weak features. 
  For the H$\delta$ line, we only used the wider window of
 4082$-$4122\AA\ as a line region to ease the comparison with models
 \citep[c.f. ][]{2003PASJ...55..771G}. 
 A histogram of H$\delta$ EW distribution is shown in Figure \ref{fig:hd_hist}.

\begin{figure}
\begin{center}
\includegraphics[scale=0.5]{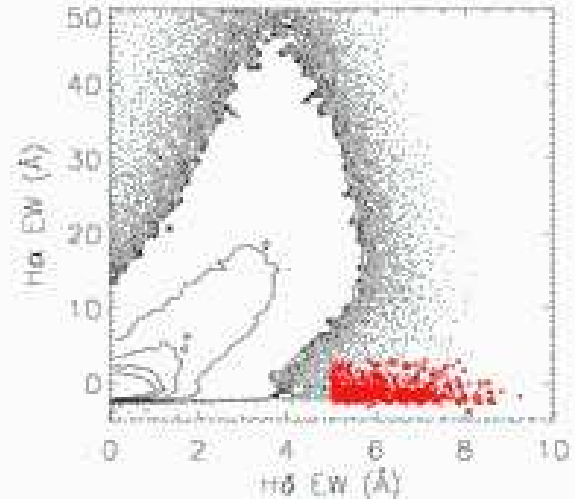}
\end{center}
\caption{ H$\alpha$ EW (emission) is plotted against  H$\delta$ EW
 (absorption) for galaxies with SN($r$)$>10$. E+A galaxies  selected in this paper  are shown with larger dots. 
}\label{fig:hd_ha}
\end{figure}

\begin{figure}
\begin{center}
\includegraphics[scale=0.5]{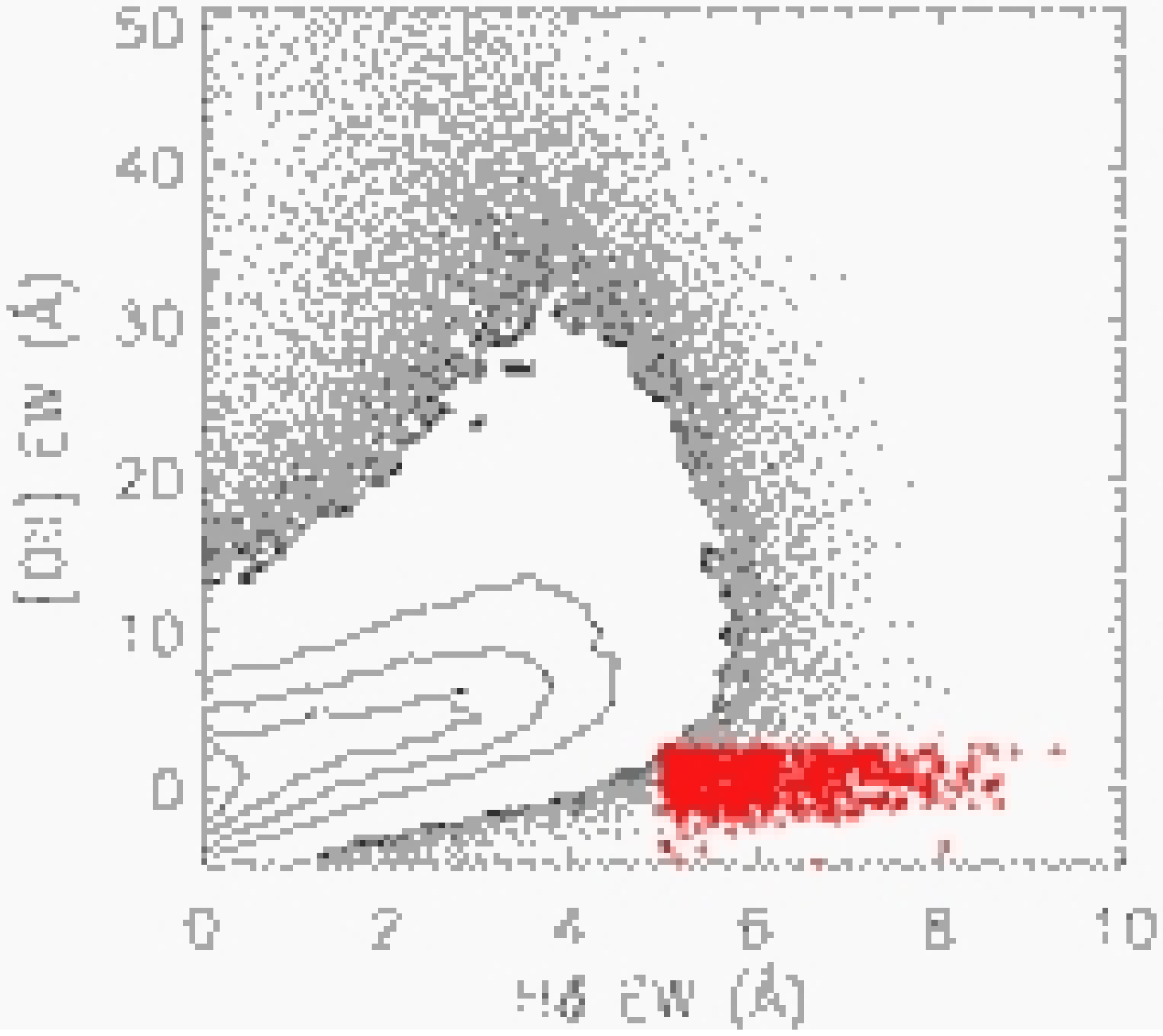}
\end{center}
\caption{ [OII] EW (emission) is plotted against  H$\delta$ EW
 (absorption). E+A galaxies  selected in this paper  are shown with larger dots. 
}\label{fig:hd_oii}
\end{figure}

  Once these lines are measured, we select E+A galaxies as those with H$\delta$~EW~$>$~5.0\AA, and H$\alpha$ EW $>$ $-$3.0\AA, and [OII] EW $>$ $-$2.5\AA\footnote{Absorption lines have a positive sign throughout this paper.}. Although the criteria on emission lines allow small amount of emission in the E+A sample, they are relatively small amount in terms of the star formation rate (SFR).   We also exclude galaxies at $0.35<z<0.37$ from our sample because of the sky feature at 5577\AA\ .
 Our criteria are more strict than previous ones (e.g., H$\delta$ EW $>$
 4.0\AA\ and [OII]~EW~$> -$2.5\AA), suppressing possible contaminations
 from other populations of galaxies \citep{2004A&A...427..125G}.
 To illustrate our selection criteria, we plot H$\alpha$ EW (emission) against  H$\delta$ EW
 (absorption) for galaxies with SN($r$)$>10$ in Figure
 \ref{fig:hd_ha}. E+A galaxies are shown in larger dots.
 In Figure \ref{fig:hd_oii}, we show [OII] EW (emission) against  H$\delta$ EW
 (absorption) for galaxies with SN($r$)$>10$.
 Compared with the distribution of all galaxies shown in the contours,
 these two figures show that E+A galaxies have a unique
 emission/absorption line properties.

We stress an advantage in using the H$\alpha$ line in selecting E+A galaxies. Previous samples of E+A galaxies were often selected based solely on [OII] emission and Balmer absorption lines either due to the high redshift of the samples or due to instrumental reasons. \citet{2003PASJ...55..771G} showed that such selections of E+A
 galaxies without information on H$\alpha$ line would suffer from 52\%
 of contamination from H$\alpha$ emitting galaxies, whose morphology and color are very different from that of E+A galaxies \citep{2003PASJ...55..771G}.
To back this up, \citet{2004MNRAS.355..713B} selected E+A galaxies from the 2dF using only Balmer and [OII] lines to find that some E+A galaxies in their sample have the H$\alpha$ line in emission. When H$\alpha$ line is not available for the E+A selection, since H$\beta$ and H$\gamma$ absorption features are subject to the emission filling, \citet{2004MNRAS.355..713B} suggested that using three Balmer absorption lines (H$\delta$, H$\gamma$ and H$\beta$) can supress the contamination from the galaxies with detectable H$\alpha$ emission.

\subsection{Catalog of 564 E+A galaxies}\label{catalog}

 Among 451759 galaxies which satisfy the redshift and S/N
  criteria with measurable [OII], H$\delta$ and H$\alpha$ lines, we have found 564 E+A galaxies. This is one of the largest sample of E+A galaxies to date. 
  The 564 E+A galaxies span a redshift range of 0.0327$\leq z \leq$0.3421, within which H$\alpha$ line is securely covered by the SDSS spectrograph. %Although this range covers about 2 gigayear of lookback time, majority of our E+A galaxies lie at $z\sim 0.1$. Therefore, the evolutionary effect within the E+A sample in this paper.

\begin{figure*}
\begin{center}
\includegraphics[scale=0.45]{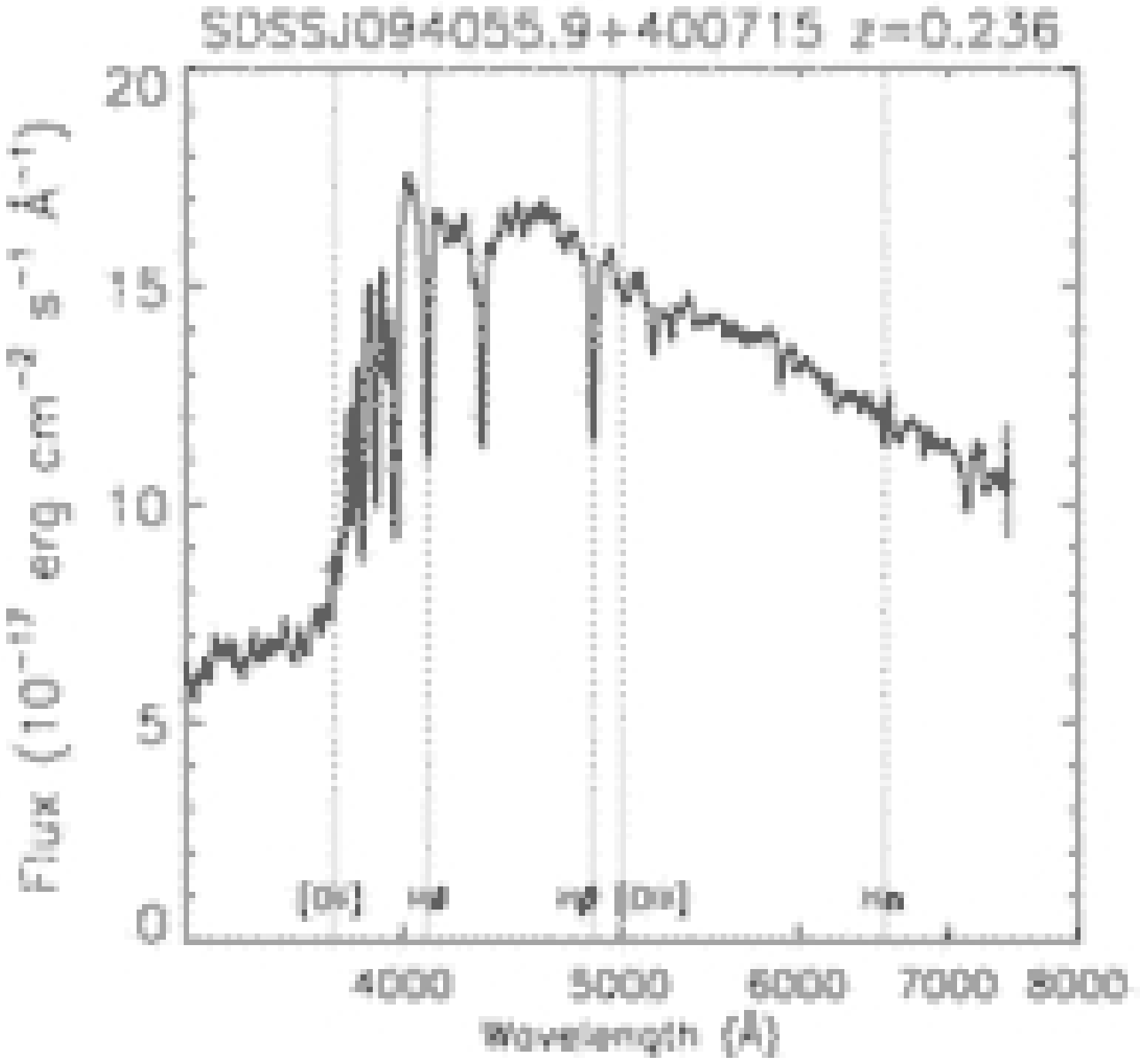}
\includegraphics[scale=0.45]{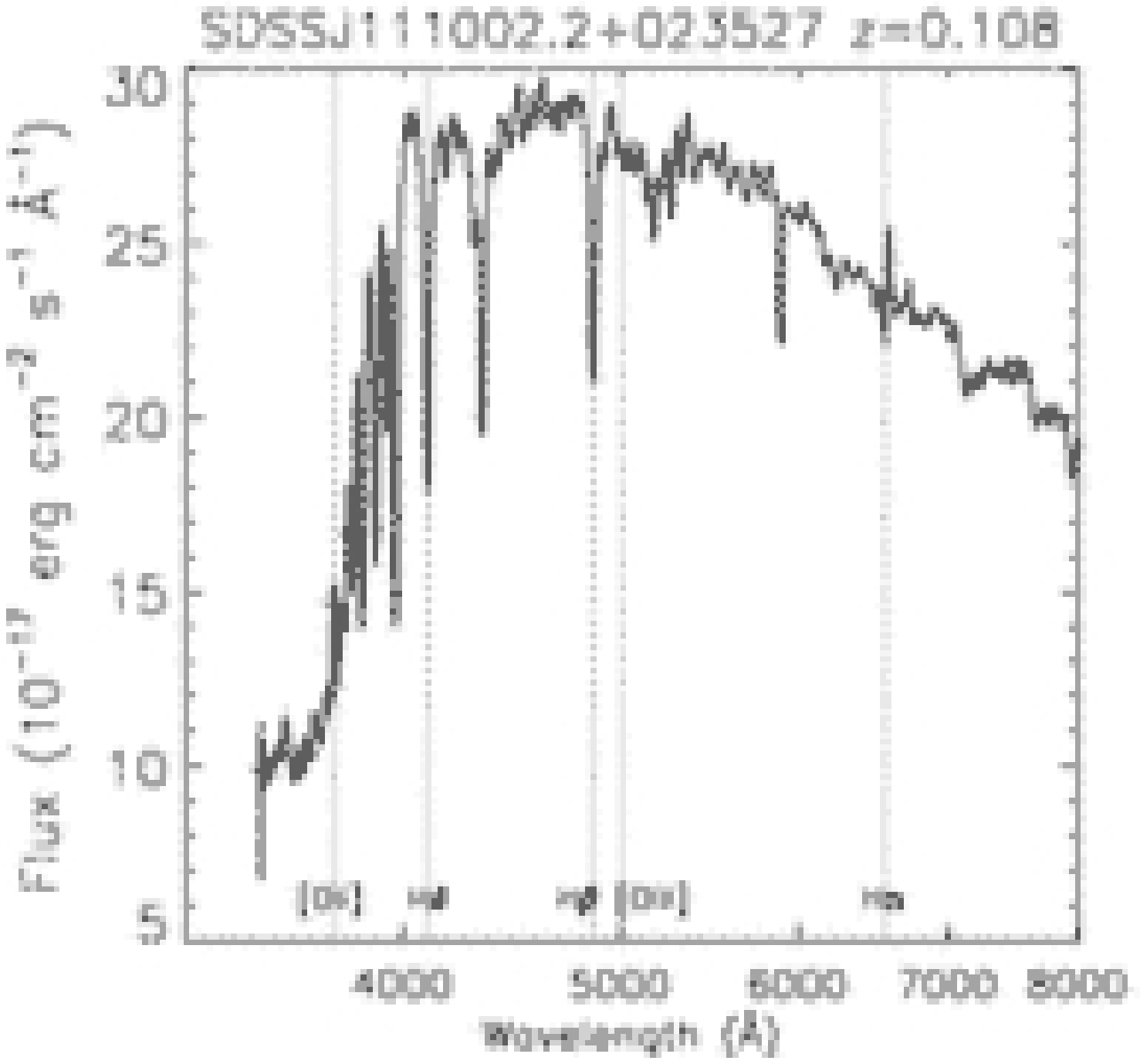}
\includegraphics[scale=0.45]{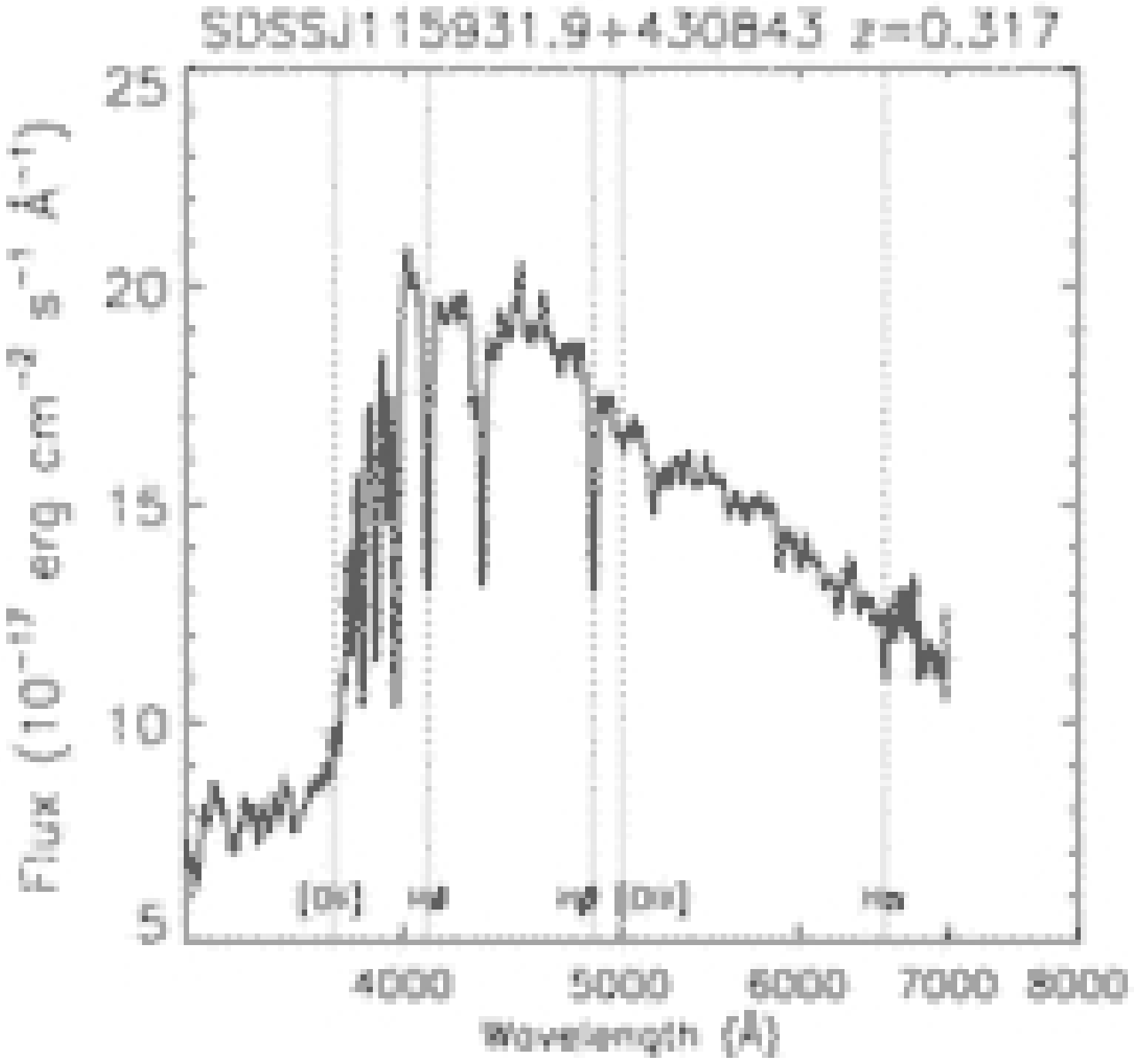}
\includegraphics[scale=0.45]{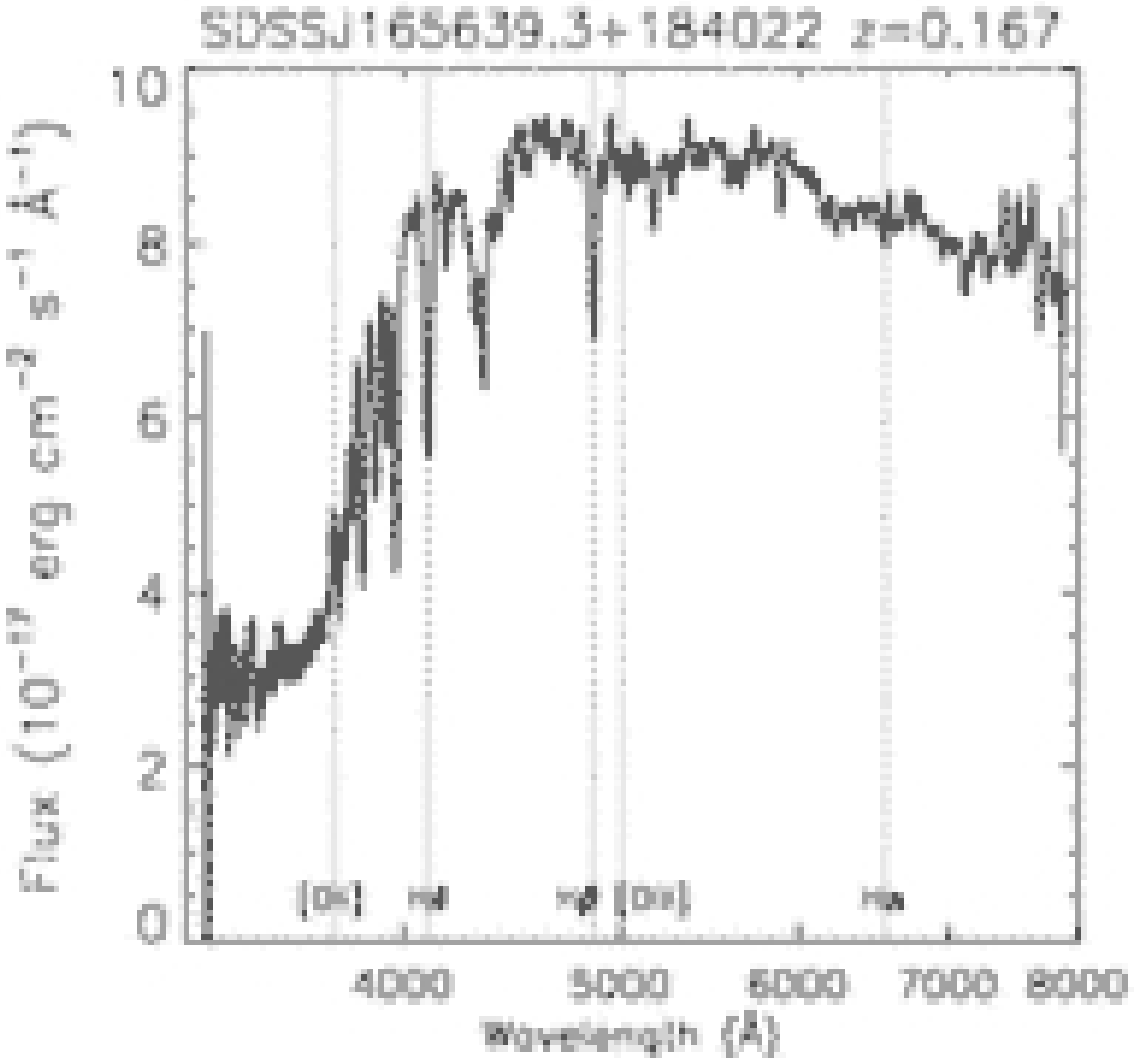}
\end{center}
\caption{Example spectra of 4 E+As with largest H$\delta$ EW of 8.62-9.27\AA.  Each spectrum is shifted to the restframe wavelength and smoothed using a 20 \AA\ box. The corresponding images are shown in Fig. \ref{fig:ea7_images}. (The image of the same galaxy can be found in the same column/row panel of Fig.\ref{fig:ea7_images})
}\label{fig:ea7_spectra}
\end{figure*}

\begin{figure*}
\begin{center}
\includegraphics[scale=0.39]{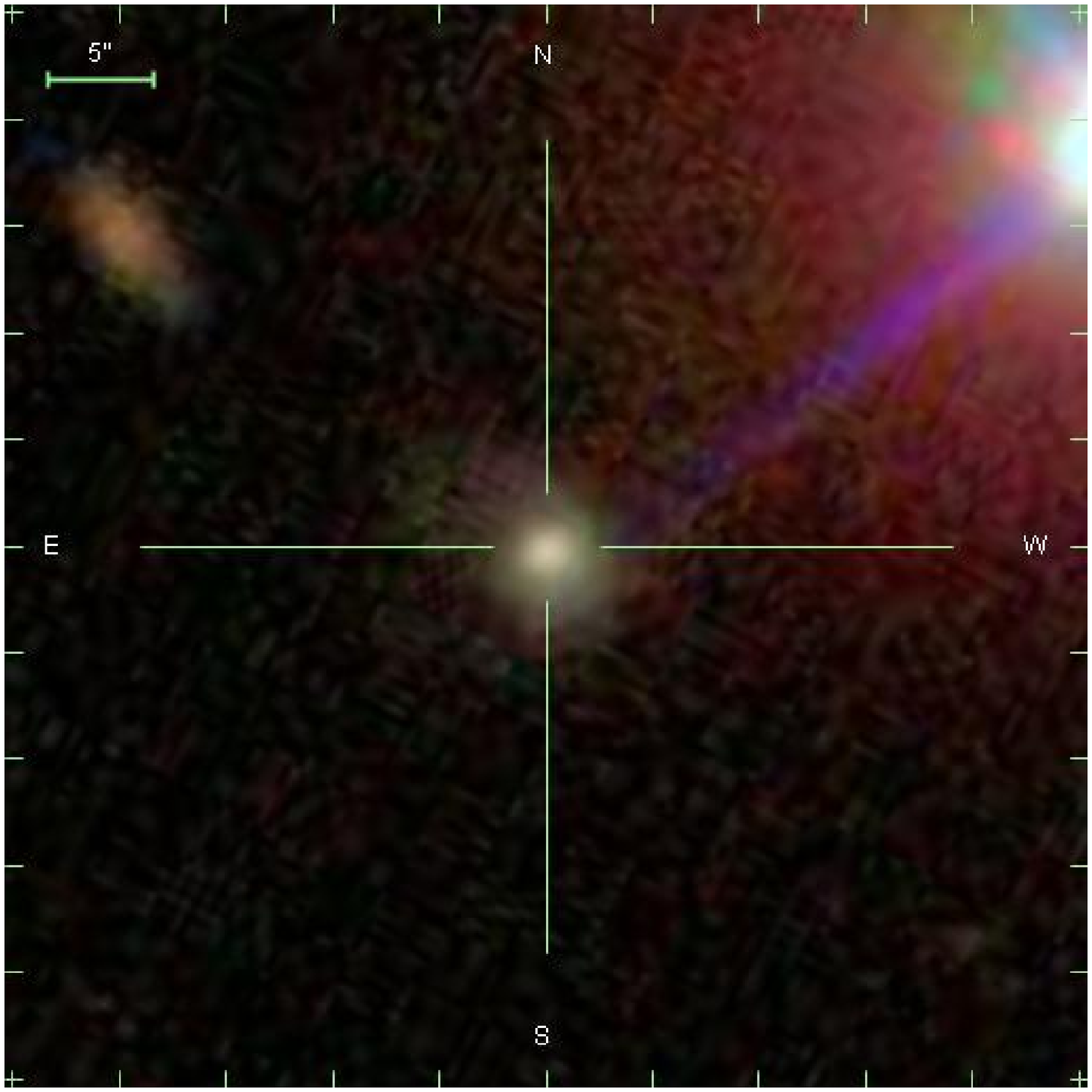}
\includegraphics[scale=0.39]{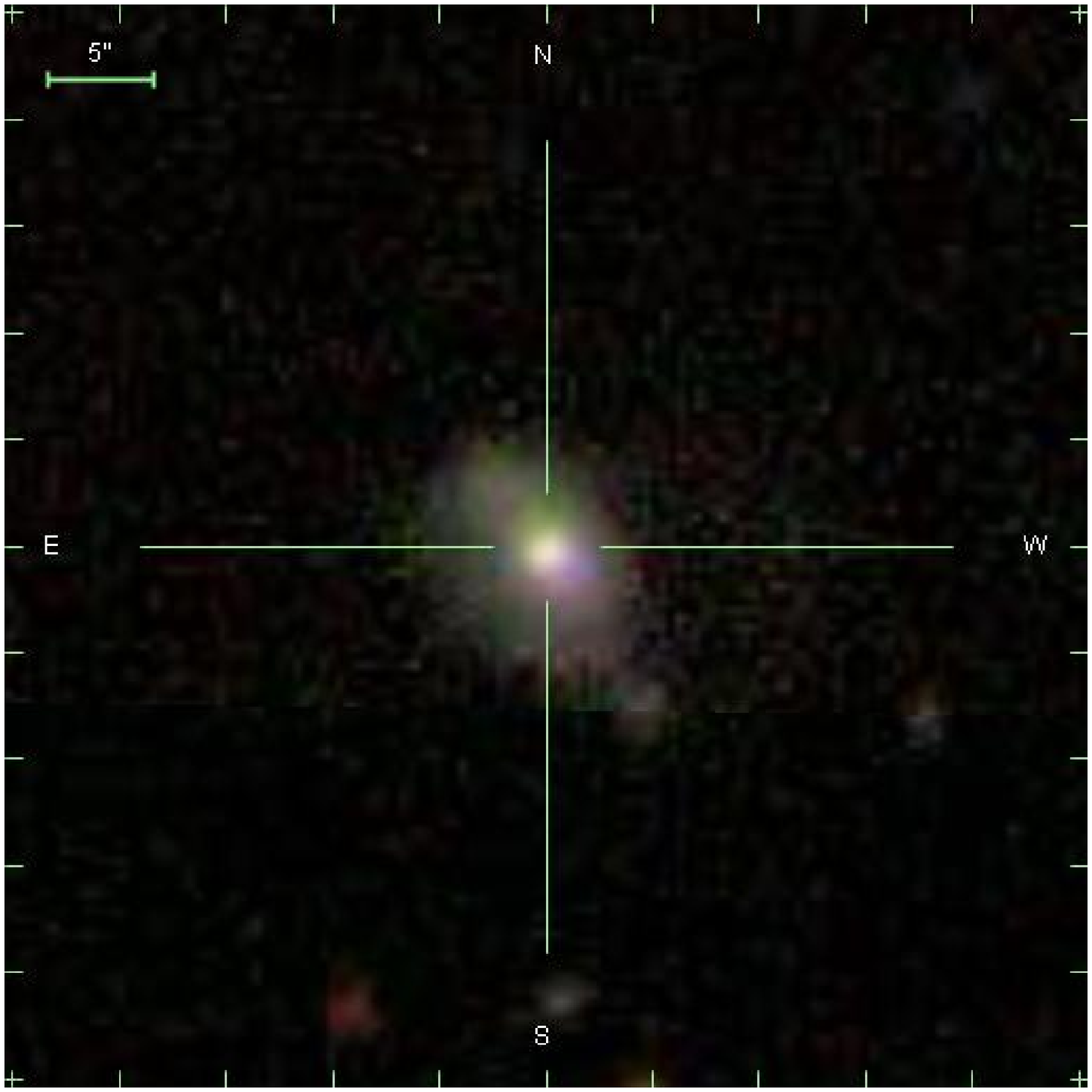}
\includegraphics[scale=0.39]{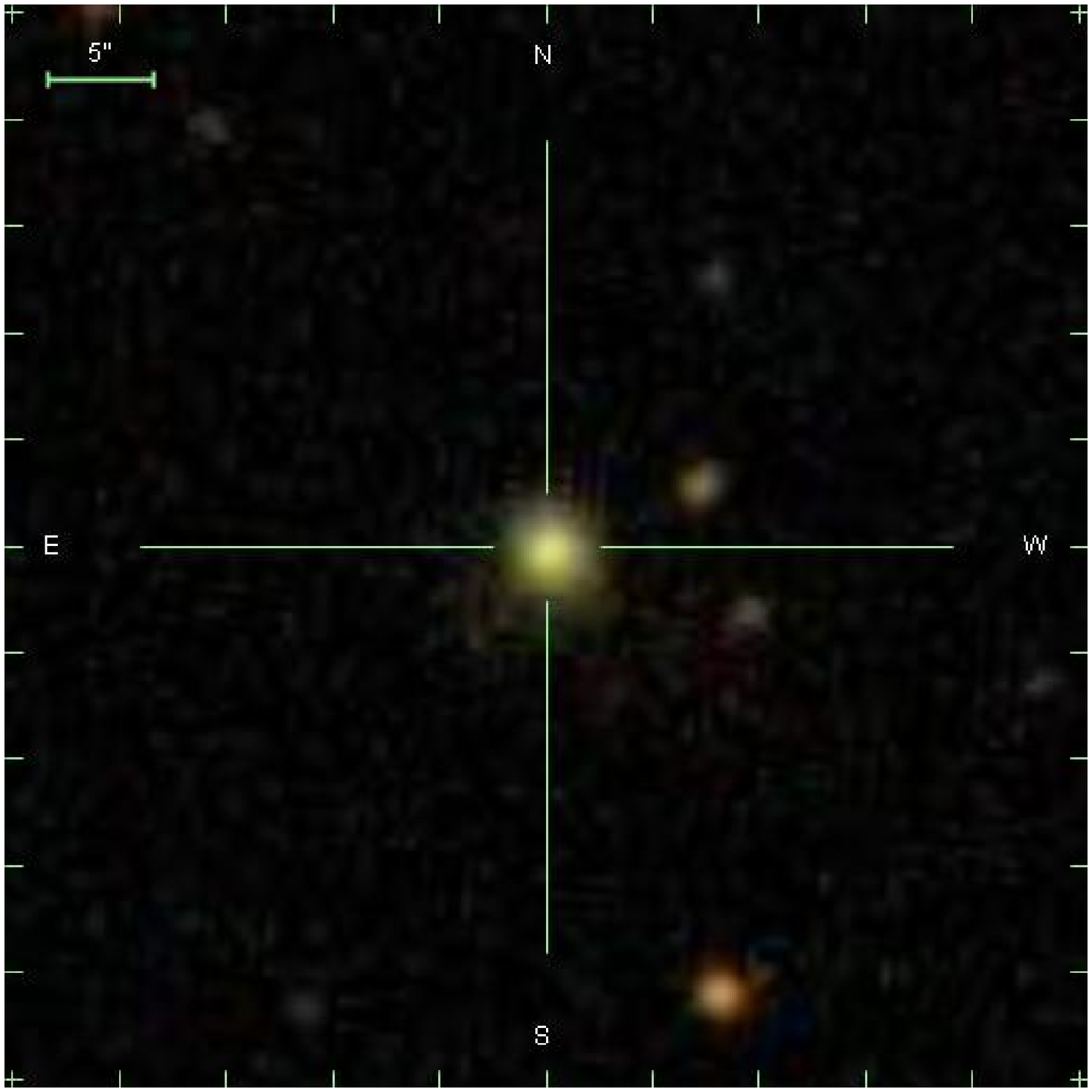}
\includegraphics[scale=0.39]{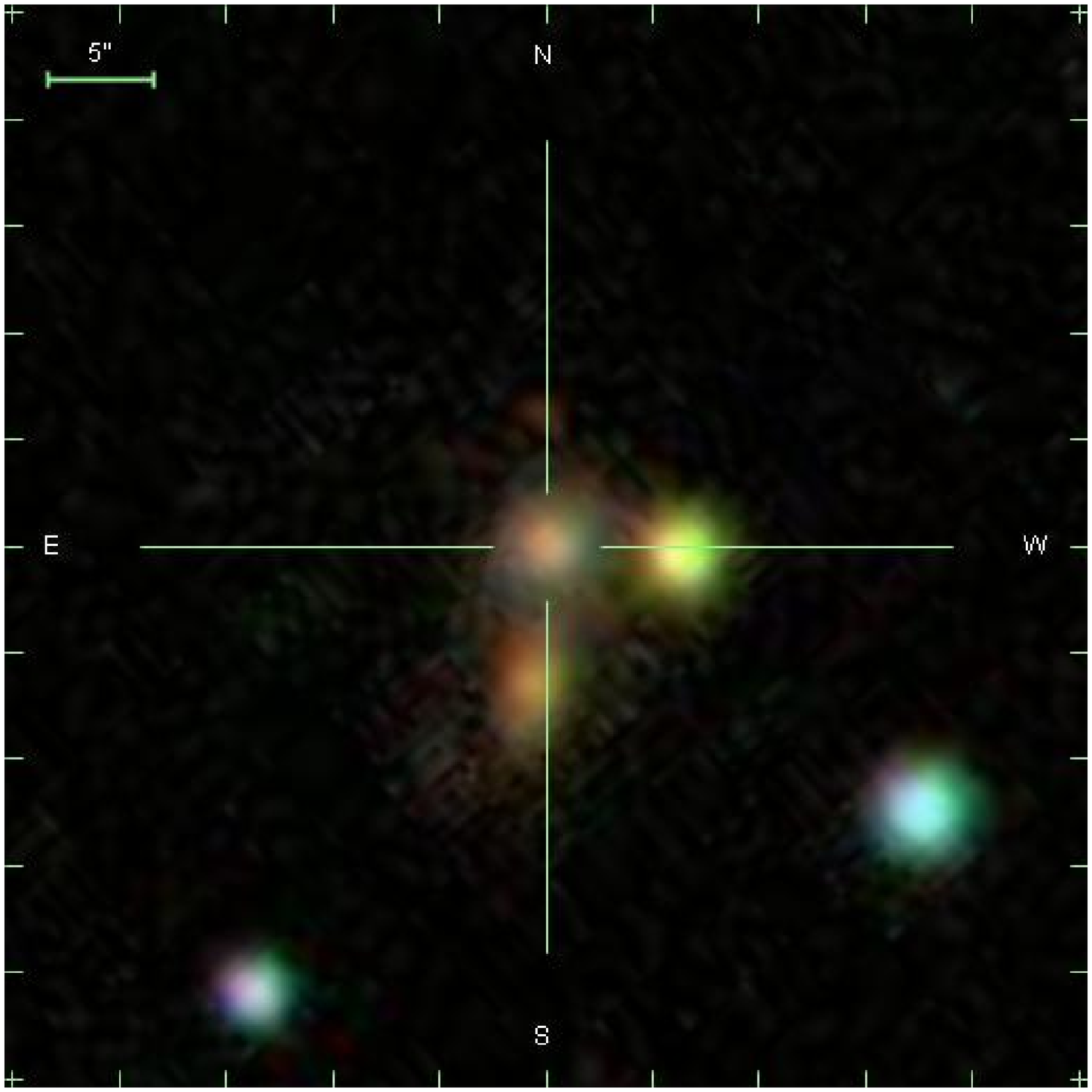}
\end{center}
\caption{Examples of $g,r,i$-composite images of E+A galaxies with 4 H$\delta$ EW $>$7\AA. The images are sorted from low to high redshift. Only 24 lowest redshift E+As are shown. The corresponding spectra with name and redshift are presented in Fig. \ref{fig:ea7_spectra}. (The spectrum of the same galaxy can be found in the same column/row panel of Fig.\ref{fig:ea7_spectra}.)
}\label{fig:ea7_images}
\end{figure*}

In Figure \ref{fig:ea7_spectra}, we show four example spectra of E+A
galaxies with strong H$\delta$ EW of 8.62-9.27\AA .  The corresponding
$g,r,i$-composite images are shown in Figure \ref{fig:ea7_images}. 

%For a comparison purpose, we also select emission line galaxies with comparable Balmer line strength of $H\delta$ EW $>$ 5.0\AA. For these galaxies, Balmer absorption lines suffer from significant amount of the emission filling, which we have corrected using the H$\alpha$/H$\beta$ ratio and the iteration procedure described in \citet{2003PASJ...55..771G}. We call these galaxies as star-forming galaxies hereafter. 

 The catalog is publicly available in the Table at  {\tt
 http://www.ir.isas.jaxa.jp/$^{\sim}$tomo/ea5} . The columns are name,
 z, RA, DEC, H$\delta$ EW, errors of H$\delta$ EW, [OII] EW, errors of
 [OII] EW, H$\alpha$ EW, errors of H$\alpha$ EW err, Petrosian magnitude in $r$, signal-to-noise ratio in $r$-band wavelength range, and a link to the NASA skyview $R$ and $K$ images.  On the object names, there is also a link to the SDSS object explorer where one can see a $g,r,i$-composite image and a GIF spectrum.

\section{Discussion}
 \subsection{Redshift Distribution}

\begin{figure}
\begin{center}
\includegraphics[scale=0.5]{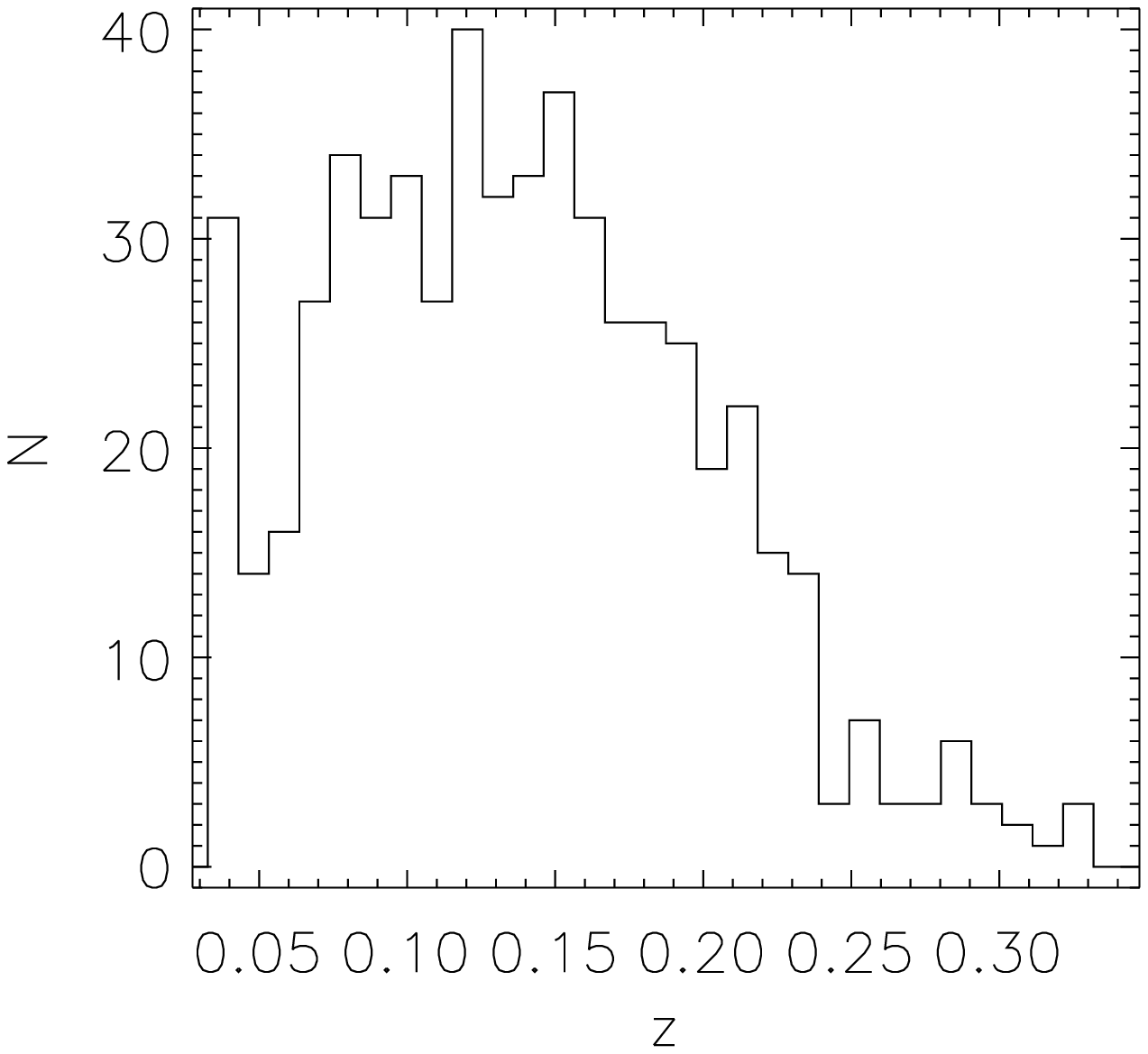}
\end{center}
\caption{
 Redshift distribution of the E+A galaxies.  
}\label{fig:zhist}
\end{figure}

\begin{figure}
\begin{center}
\includegraphics[scale=0.5]{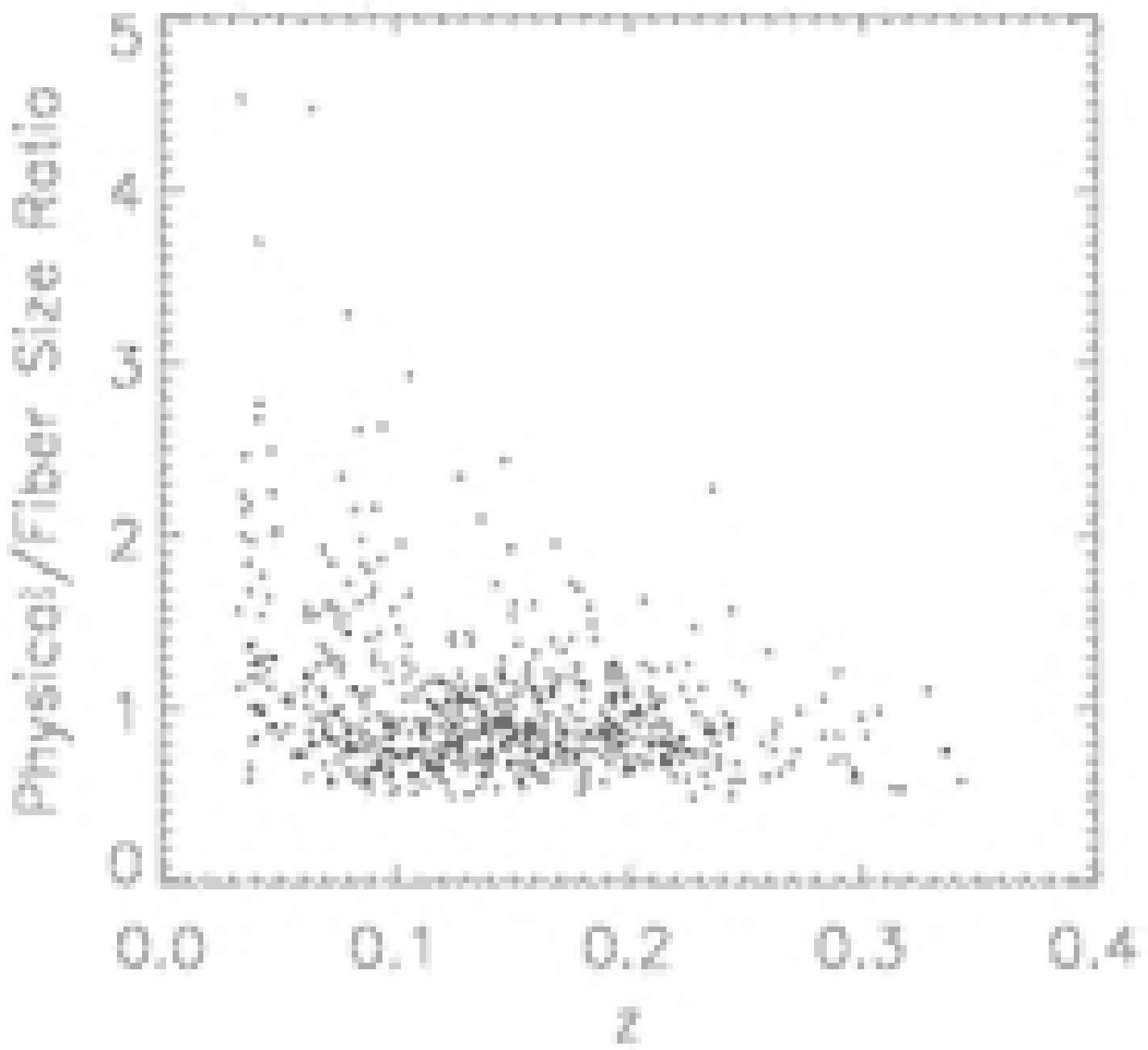}
\end{center}
\caption{
The ratio of the physical galaxy size (taken from the petrosian 50\%
 light radius) to the SDSS fiber size is shown as a function of redshift for the E+A galaxies.
}\label{fig:070124_fiber_physical_ratio.ps}
\end{figure}

 In Figure \ref{fig:zhist}, we show the redshift distribution of the E+A
 galaxies. The median redshift is z=0.138. 
 
 There is a possible peak at the lowest
 redshift bin ($z\sim 0.04$). The reason for the peak is not
 certain. 
 We have checked all the images of $z<0.05$ E+A galaxies by
 eye to find that all of them show strong H$\delta$ absorption,
 confirming that our selection criteria work properly. 
We, however, suspect the aperture bias may be one of the
 causes; the SDSS fiber spectrograph only samples the
light within the inner 3'' \citep{2002AJ....124.1810S}. This does not bring a large bias into the
distant E+A galaxies, whose sizes are smaller (the median Petrosian 50\%
 radius of our sample is 1.4''), but it does bring a
large aperture bias to nearby E+A galaxies due to their large apparent
size. At $z\sim 0.04$, the median physical size (taken from the
 Petrosian 50\% light radius) of E+A galaxies is 2.0'', whereas the
 3''-fiber only collects the light from the inner 1.5'' of radius.
  In Figure \ref{fig:070124_fiber_physical_ratio.ps}, we plot the ratio
 of Petrosian 50\% light radius to the fiber radius (1.5'') as a
 function of redshift. The figure
 shows that at $z\sim 0.04$, there are significant number of E+A
 galaxies a few times larger than the fiber size. 
  Therefore, we recommend to use a low redshift cut \citep[e.g., $z>0.05$; see
 also ][]{2003ApJ...584..210G,2003MNRAS.346..601G}
 when one performs statistical analysis on the sample.

 These nearby E+A galaxies may have remaining star-formation activity
 outside of the 3'' fiber. Nevertheless, it is also true that these
 galaxies have a post-starburst stellar population within the 3'' of
 radius. It is still important to investigate what caused the
 central post-starburst in these galaxies, and thus, we keep these
 nearby E+A galaxies in our sample. Due to their larger apparent size,
 these E+A galaxies are suitable targets for spatially-resolved
 observations \citep[e.g., ][]{2006ApJ...642..152Y,2006AJ....131.2050Y}.

\begin{figure*}
\begin{center}
\includegraphics[scale=0.45]{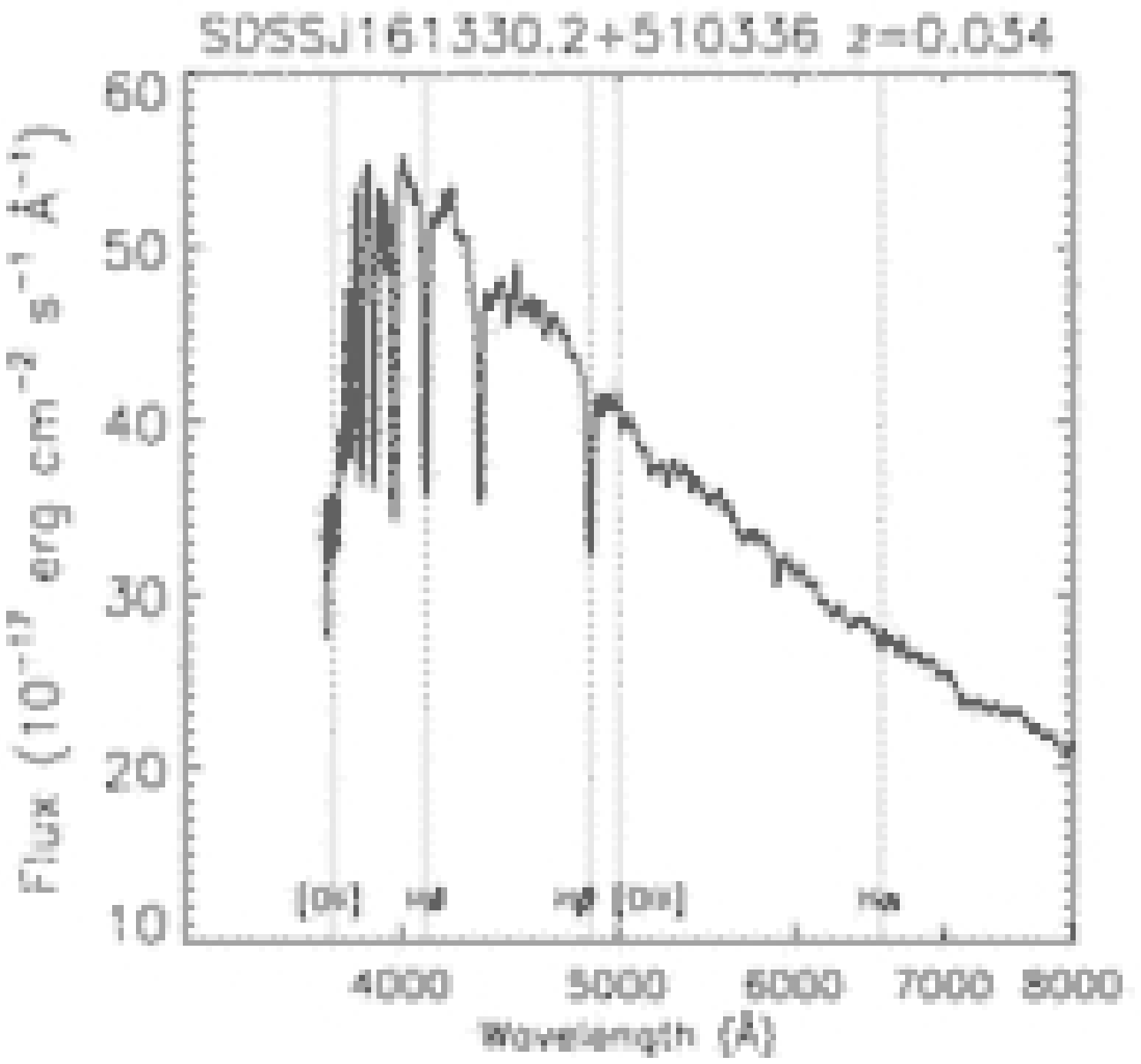}
\includegraphics[scale=0.45]{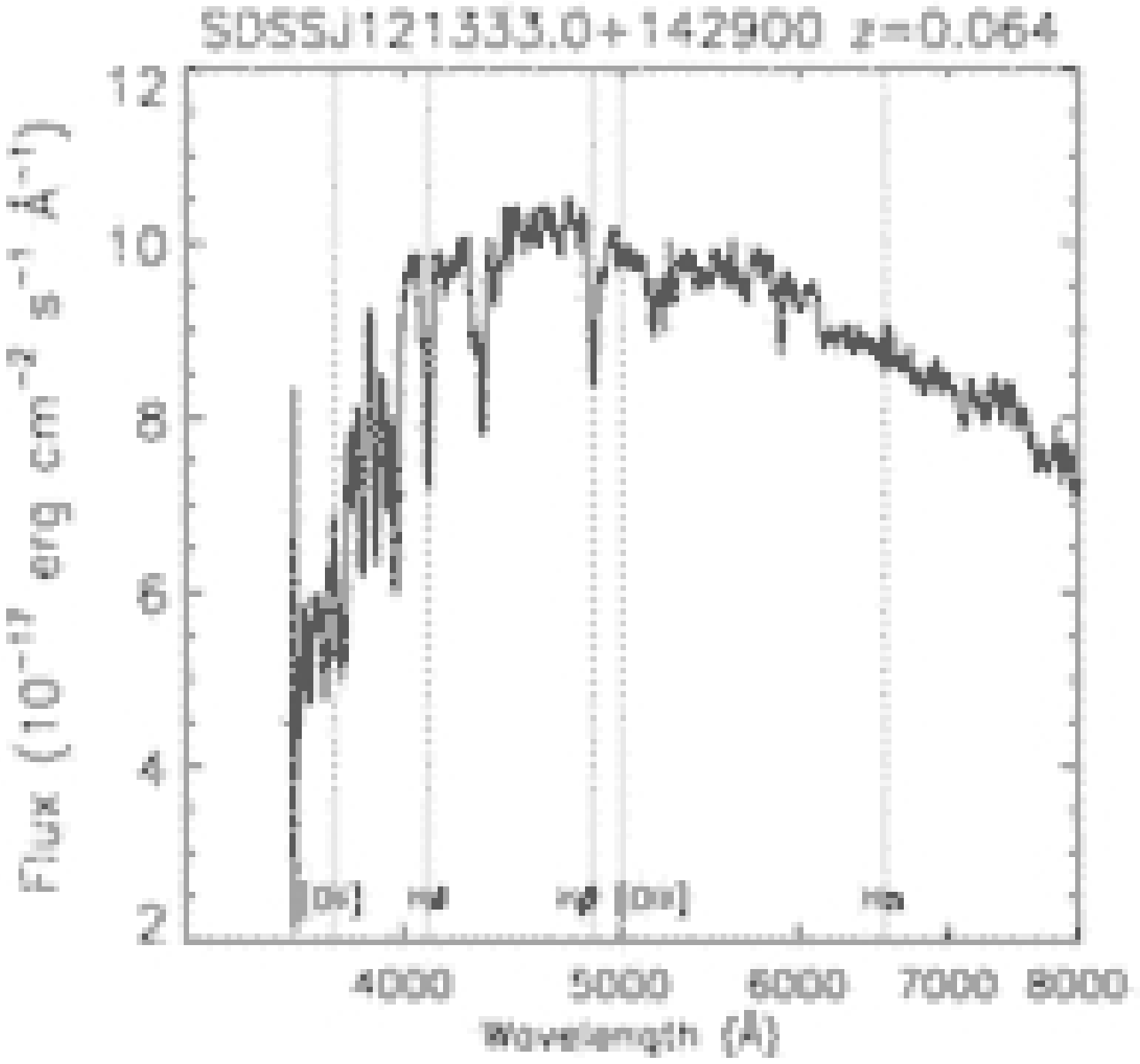}
\includegraphics[scale=0.45]{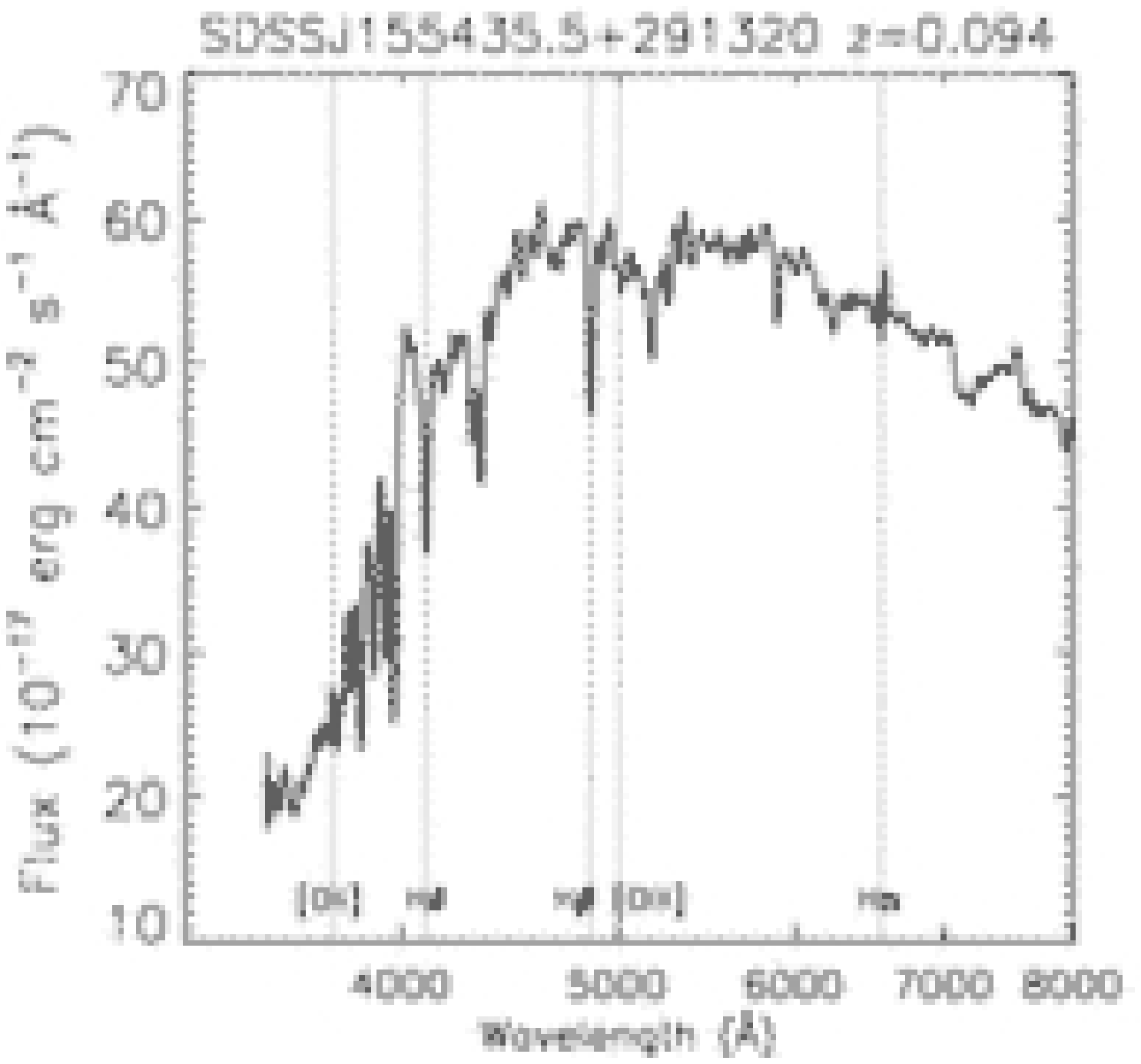} 
\includegraphics[scale=0.45]{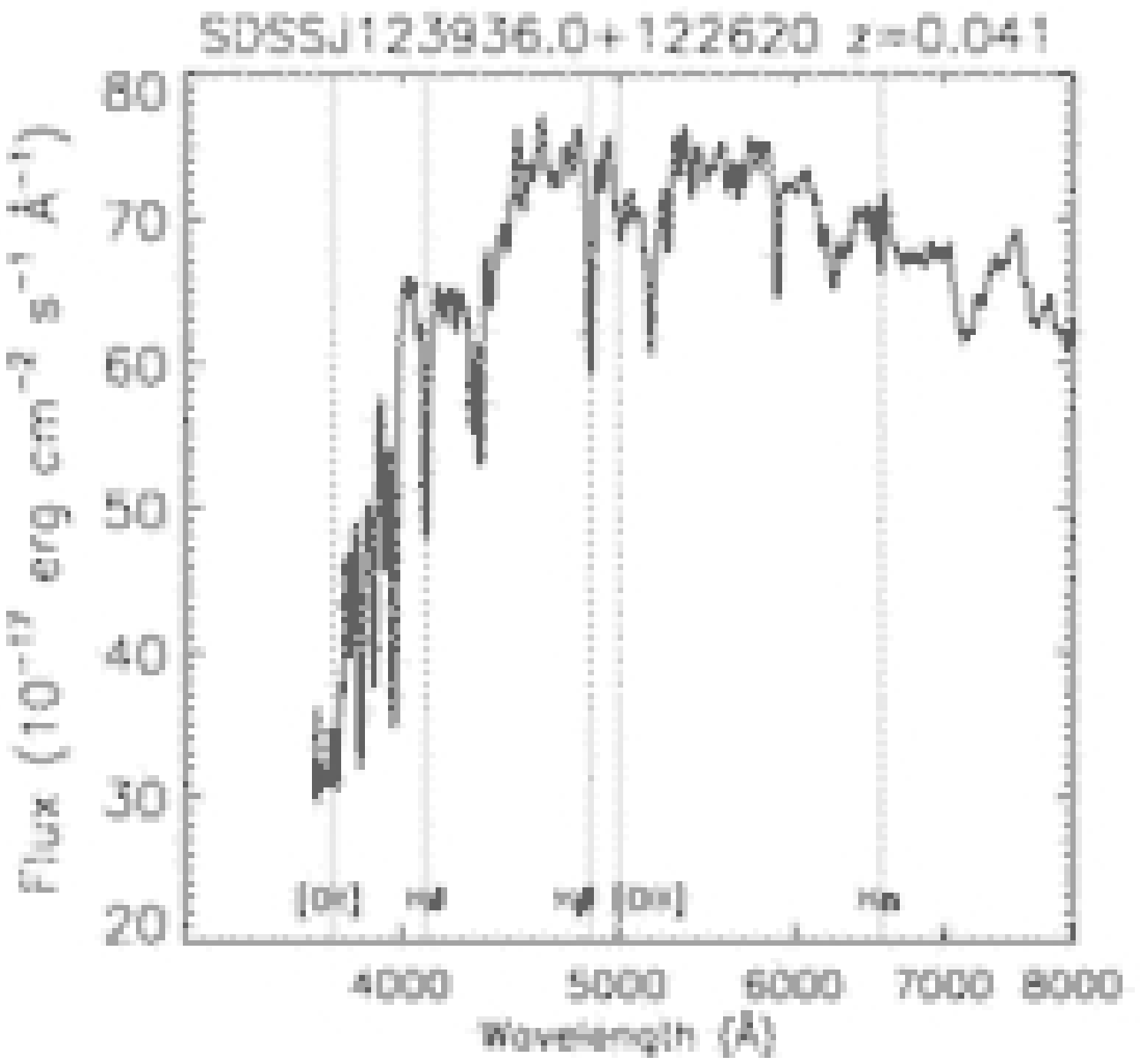}
\end{center}
\caption{Example spectra of 4 E+As with largest Petrosian 90\% light
 radii ($>14''$).  Each spectrum is shifted to the restframe wavelength and smoothed using a 20 \AA\ box. The corresponding images are shown in Fig. \ref{fig:rad_images}. (The image of the same galaxy can be found in the same column/row panel of Fig.\ref{fig:rad_images})
}\label{fig:rad_spectra}
\end{figure*}

\begin{figure*}
\begin{center}
\includegraphics[scale=0.39]{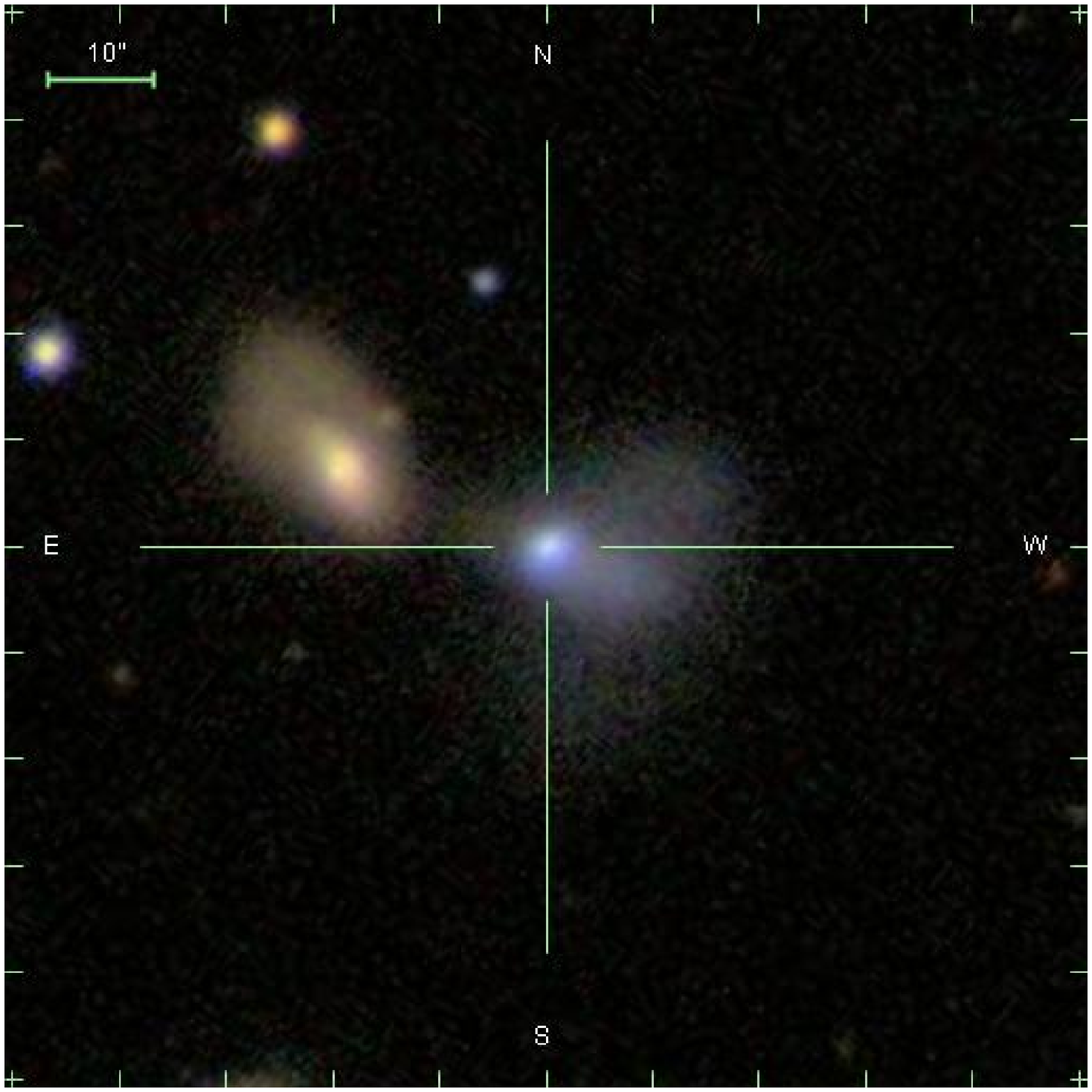}
\includegraphics[scale=0.39]{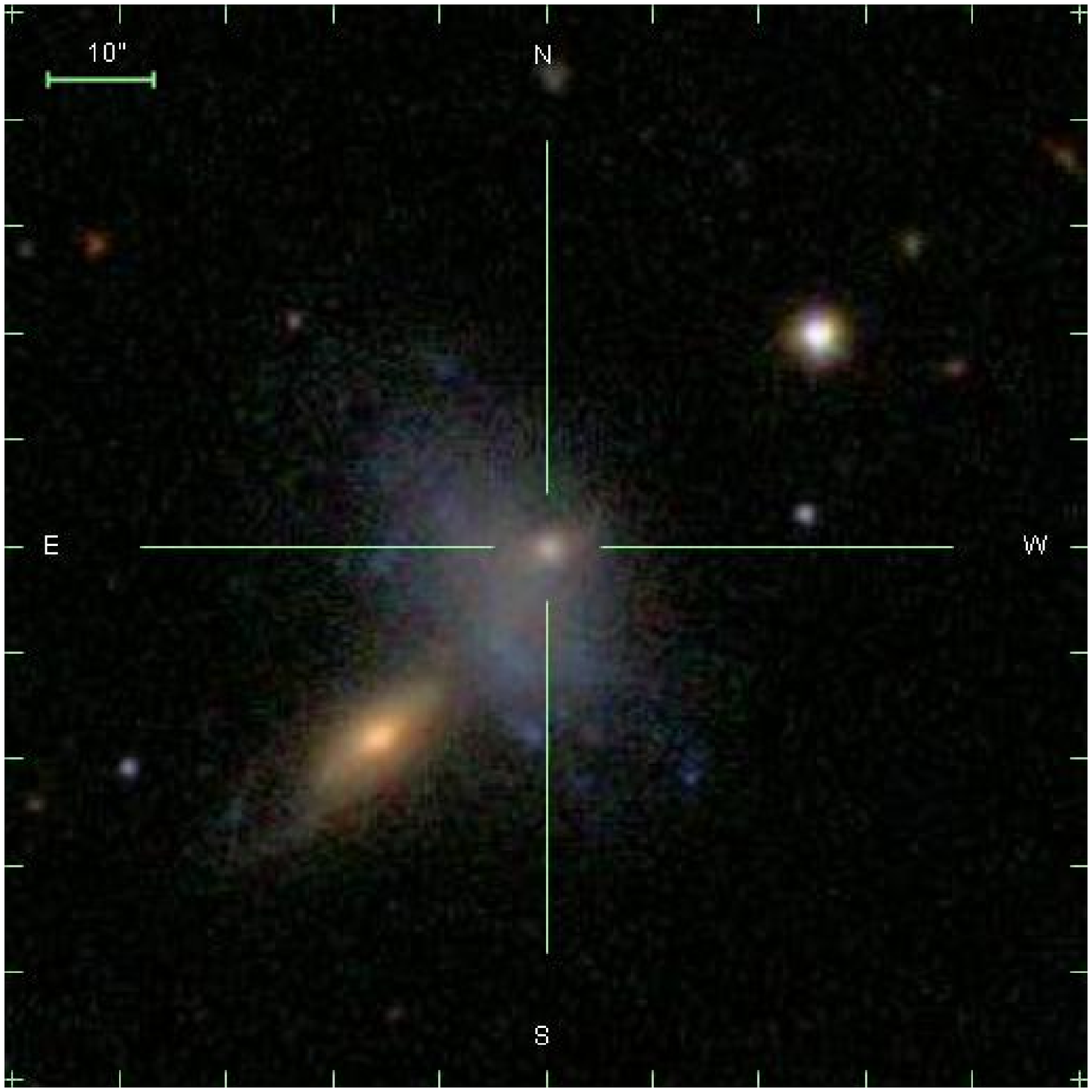}
\includegraphics[scale=0.39]{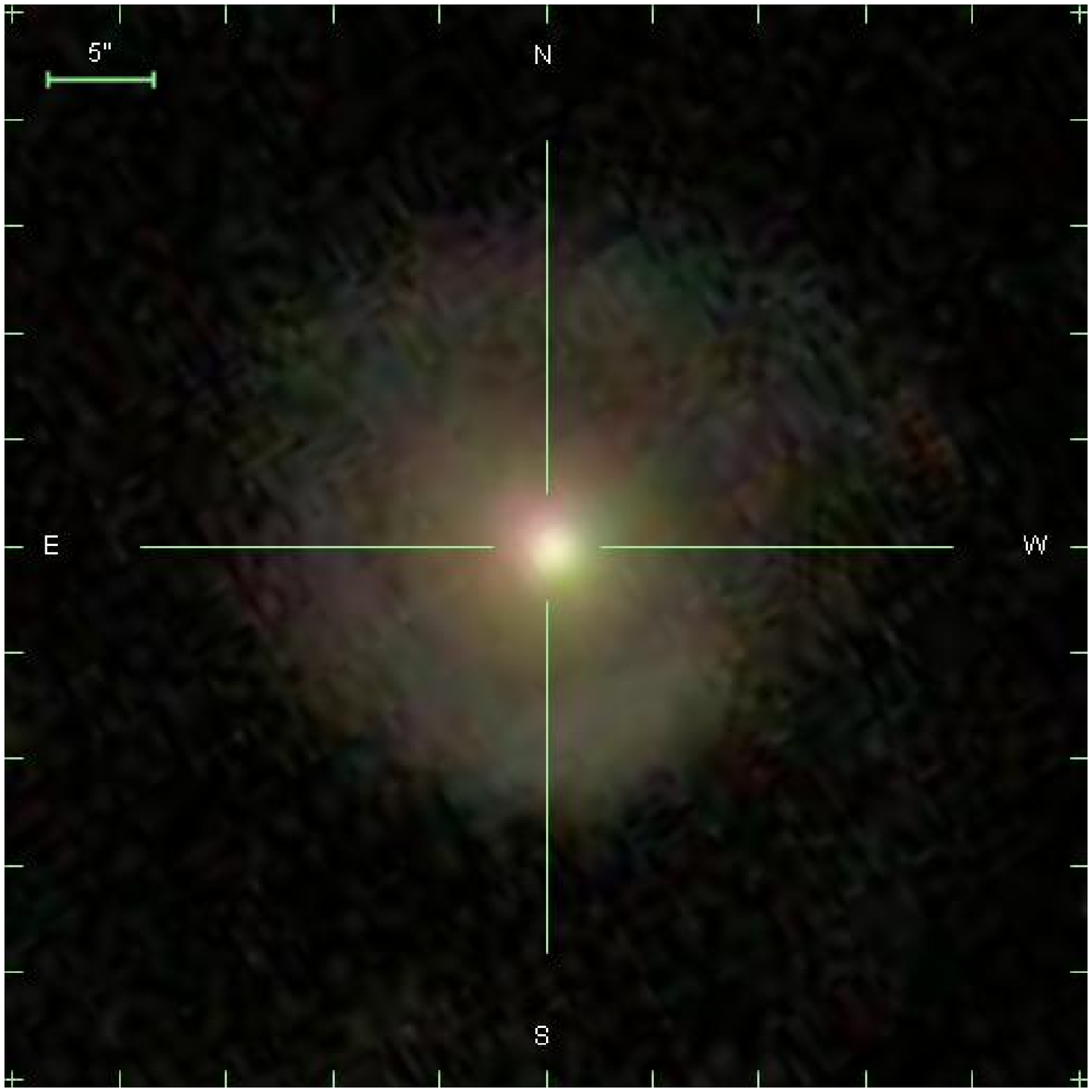}
\includegraphics[scale=0.39]{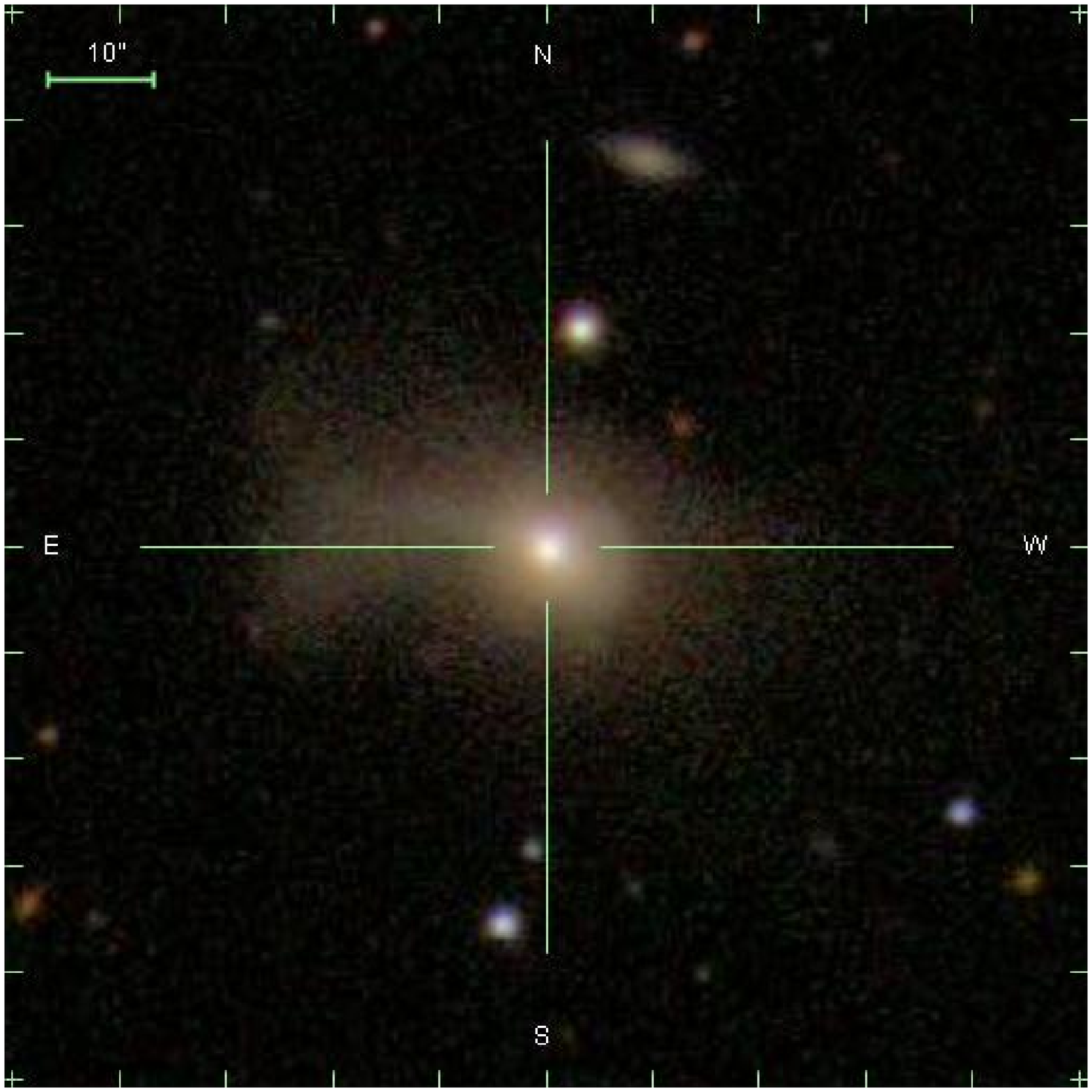}
\end{center}
\caption{Examples of $g,r,i$-composite images of E+A galaxies with
 largest Petrosian 90\% light radii ($>14''$). The corresponding spectra with name and redshift are presented in Fig. \ref{fig:rad_spectra}. (The spectrum of the same galaxy can be found in the same column/row panel of Fig.\ref{fig:rad_spectra})
}\label{fig:rad_images}
\end{figure*}

In Figure \ref{fig:rad_spectra}, we show four example spectra of E+A
galaxies with a large apparent size of Petrosian 90\% radius of $>14''$. 
 These E+A galaxies are large enough for detailed morphological/substructure studies.  The corresponding $g,r,i$-composite images are shown in Figure \ref{fig:rad_images}. There is a hint of possible dynamical disturbances in all four galaxies shown here.

% \subsection{Aperture bias}
%Yagi et al. 
%We caution readers that 

 \subsection{E+As in the high redshift Universe}

Our catalog of E+A galaxies provides us with a useful benchmark in
studying the evolution of E+A galaxies in higher redshift Universe.
%Assuming that the lifetime of the E+A phase is $\sim$1 gigayear based on
%the lifetime of A-type stars, only 2-3\% of galaxies have been through
%the E+A phase in the Universe of 13.7 gigayears of age, if the fraction
%of the E+A galaxies has been constant at $\sim$0.2 \%. This, however, is
%a too simple assumption. 
 It has been known that cosmic star formation density was higher at
 $z>1$ \citep{1996MNRAS.283.1388M}. According to
 \citet{2006ApJ...642...48L}, the fractions of the post-starburst
 galaxies was larger by a factor of 2 at $z\sim 1$ \citep[also see ][]{2006MNRAS.373..349R}.  By comparing the fraction of E+A galaxies in different redshift, we can trace the cosmic star formation history in terms of the post-starburst galaxies.

%If we loosen our criteria to select post-starburst galaxies to H$\delta$\_EW $>$3\AA, the fraction increase by a factor of $\sim$3 \citep{2003PASJ...55..771G}. Therefore, an estimate on the high side yield $\sim$15\% of galaxies may have been through the E+A phase. However, this is still a small fraction of galaxies compared to the total number of galaxies and suggests that not all the elliptical galaxies have not been through the E+A phase. 
% Therefore, it is likely that E+A galaxies are only one of the multiple progenitors of elliptical galaxies.
% If star-forming galaxies decrease their star formation rate more slowly, for example, exponentially with $\tau\sim$1 gigayear, the galaxies do not experience the E+A phase before they become passive galaxies \citep{2003PhDT.........2G}. 
%It is important to investigate what special physical mechanism is needed for a galaxy to become an E+A galaxy. 

 In high redshift cluster environments, E+A galaxies are much more numerous; 
 pioneering work were done by
 \citet{1999ApJS..122...51D,1999ApJ...518..576P}, who  found E+A galaxies (H$\delta$ EW$>$3\AA and not detectable emission in [OII]) are significantly more common in 10 clusters at 0.37$<z<$0.56 than in the field (21$\pm$2\% compared with 6$\pm$3\%).
 Later, \citet{2003ApJ...599..865T}  found 7-13\% of E+A galaxies in three high redshift clusters at z=0.33,0.58, and 0.83, claiming that $>30$\% of E+S0 members may have undergone the E+A phase if the effects of E+A downsizing and increasing E+A fraction as a function of redshift are considered (their selection criteria is $\frac{H\delta EW+H\gamma EW}{2} >4\AA$ and  [OII] EW $>$-5\AA ). In their search for field E+A galaxies among 800 spectra, \citet{2004ApJ...609..683T}   measured the E+A fraction at 0.3 $< z <$ 1 to be 2.7\% $\pm$ 1.1\%, a value lower than that in galaxy clusters at comparable redshifts.
  \footnote{Note that  \citet{1999ApJ...527...54B} found smaller E+A fractions of  1.5\% $\pm$ 0.8\% in high redshift clusters ($z\sim$0.25).  However, it has been controversial whether there is a discrepany on the fraction of E+A galaxies in the high redshift clusters. See Section 5.1 of \citet{2004ApJ...617..867D} for detailed discussion.}

% Note that there is an exception; Based on 1823 galaxies in the 15 X-ray
% luminous clusters at 0.18 $< z <$ 0.55, on the contrary,
% \citet{1999ApJ...527...54B} found that the fraction of K+A galaxies is
% 1.5\% $\pm$ 0.8\% in the cluster and 1.2\% $\pm$ 0.8\% in the field,
% i.e., they found no difference. 
% However, it is not clear if there is a discrepany exists on the fraction of E+A galaxies in the high redshift cluster environment.
% Possible causes include difference in cluster sample (differt X-ray luminosity and different redshift), different selection criteria (H$\delta$ EW of 3 and 5A\\ ), and differnt method used in measuring H$\delta$ EW. See Section 5.1 of \citet{2004ApJ...617..867D} for more detailed discussion on the possible causes in the difference of E+A fractions in the higher redshift cluster
% environment.

 Taken together with the much smaller fraction of E+A galaxies in the local field, the sudden increase of the fractions in distant clusters over all other environments suggests that there is something unique to this time and the environment that gives rise to so many E+A galaxies. 
It is also important to note that E+A galaxies may have heterogeneous origins; in the local Universe, E+As are preferentially found in the rarefied field regions, whereas at higher redshift, E+As are found in the cluster environment. Considering the large difference in time and the environment, E+As in the local field and high-z clusters may have been created by different physical mechanisms. 
 For example, \citet{2005MNRAS.357..937G} has shown that local (z$\sim$0.1) E+A galaxies have more close companion galaxies than average galaxies, showing that the dynamical merger/interaction could be the physical origin of E+A galaxies. However, most E+A galaxies in the high redshift clusters are known not to be major mergers \citep{1999ApJS..122...51D}.
   %It is important to create a large sample of E+A galaxies at higher redshift spanning a wide range of environment to investigate the evolution of E+A galaxies and its effect on the overall galaxy evolution in the Universe.

%Both of these work did not use H$\alpha$ line in the selection criteria, and thus, could suffer from the up to 52\% of the contamination \citep{2003PASJ...55..771G}. More statistically significant analysis based on a larger sample is needed to shed light on the galaxy evolution in the high redshift cluster environment. 
% In the low redshift cluster environment, \citep{2003PASJ...55..757G} identified that passive spiral galaxies, i.e.,  galaxies with no star formation but with spiral morphology, preferentially reside in the cluster infalling regions. These passive spiral galaxies may be a more important smoking-gun in galaxy evolution in cluster environment.

\subsection{Selection criteria of E+A galaxies}

We used the selection criteria of H$\delta$EW $>5\AA$ and H$\alpha$ EW
  $<-3.0\AA$,[OII] EW$<-2.5\AA$ to select 564 E+A galaxies (Section
  \ref{catalog}). However, we caution readers that the number (or
  fraction) of the selected E+A galaxies strongly depends on the
  selection criteria. For example, if we loosen our criteria to
  H$\delta$ EW$>4\AA$, the number of E+A galaxies increases to 1062 (almost twice as numerous). With
  H$\delta$ EW$>3\AA$, 2298 E+A galaxies are selected (4 times more numerous). 
  Therefore, when one compares samples of E+A galaxies selected from different data sets, it is important to synchronize the selection criteria. Especially, one compares to a sample at higher redshift, inconsistency in selection criteria can produce spurious  evolutionary effect.

\section{Summary}

We have constructed one of the largest catalogs of 564 local E+A (post-starburst) galaxies carefully selected from the SDSS DR5. The sample provides us with a useful tool to investigate statistical properties and/or for follow-up observations of this rare, but important population of galaxies.

\section*{Acknowledgments}

We are grateful to the referee, Prof.A.Dressler, for many insightful
comments that improved the paper significantly.
We thank Dr. M.Yagi, and C.Yamauchi for useful discussion.
%, and Dr. T.Hattori for friendly help during the observation. We thank the anonymous referee for many insightful comments, which significantly improved the paper.
%Dr H. C. Bhatt for a critical reading of the original version of the
%paper and 

% The United Kingdom Infrared Telescope is operated by the Joint Astronomy Centre on behalf of the U.K. Particle Physics and Astronomy Research Council.

% Use of the UKIRT 3.8-m telescope for the observations is supported by NAOJ.
 The research was financially supported by the Sasakawa Scientific Research Grant from The Japan Science Society.

 This research was partially supported by the Japan Society for the Promotion of Science through Grant-in-Aid for Scientific Research 18840047.
% This research is in part based on observations obtained with the Apache Point Observatory 3.5-meter telescope, which is owned and operated by the Astrophysical Research Consortium.
   
    Funding for the creation and distribution of the SDSS Archive has been provided by the Alfred P. Sloan Foundation, the Participating Institutions, the National Aeronautics and Space Administration, the National Science Foundation, the U.S. Department of Energy, the Japanese Monbukagakusho, and the Max Planck Society. The SDSS Web site is http://www.sdss.org/.

    The SDSS is managed by the Astrophysical Research Consortium (ARC) for the Participating Institutions. The Participating Institutions are The University of Chicago, Fermilab, the Institute for Advanced Study, the Japan Participation Group, The Johns Hopkins University, Los Alamos National Laboratory, the Max-Planck-Institute for Astronomy (MPIA), the Max-Planck-Institute for Astrophysics (MPA), New Mexico State University, University of Pittsburgh, Princeton University, the United States Naval Observatory, and the University of Washington.

%\appendix

%\bsp

\label{lastpage}

\end{document}